\documentclass[useAMS,usenatbib]{mn2e}
\usepackage{times}
\usepackage{graphicx}
\usepackage{amssymb}

\title{Mass estimation in the outer non-equilibrium region of galaxy clusters}
\author[Guido Cupani, Marino Mezzetti and Fabio Mardirossian]
{Guido Cupani,$^1$\thanks{E-mail: \texttt{cupani@oats.inaf.it}.}
Marino Mezzetti$^{1,2}$
and Fabio Mardirossian$^{1,2}$\\
$^1$ Dipartimento di Astronomia, Universit\`a degli studi di Trieste, via Tiepolo 11, I-34143 Trieste, Italy\\
$^2$ INAF - Istituto Nazionale di Astrofisica, via Tiepolo 11, I-34143 Trieste, Italy}
\pagerange{\pageref{firstpage}--\pageref{lastpage}}
\pubyear{2007}

\newcommand{\bi}{\begin{itemize}}
\newcommand{\be}{\begin{equation}}
\newcommand{\ei}{\end{itemize}}
\newcommand{\ee}{\end{equation}}
\newcommand{\f}{\frac}
\newcommand{\p}{\item}

\begin{document}

\maketitle

\label{firstpage}

\begin{abstract}
We discuss a new criterion to estimate the mass in the outer, non-equilibrium region of galaxy clusters, where the galaxy dynamics is dominated by an overall infall motion towards the cluster centre. In the framework of the spherical infall model the local mean velocity of the infalling galaxies at every radius provides information about the integrated matter overdensity $\delta$. Thus, a well-defined value of the  overdensity $\delta_t$ is expected at the turnaround radius $r_t$, i.e. the radius where the Hubble flow balances the infall motion. Within this scenario, we analysed the kinematical properties of a large catalogue of simulated clusters, using both dark matter particles and member galaxies as tracer of the infall motion. We also compared the simulation with analytical calculation performed in the spherical infall approximation, to analyze the dependence of the results on cosmology in spatially flat universe. If we normalize cluster mass profiles by means of the turnaround mass $M_t$ (i.e. the mass within $r_t$), they are consistent with an exponential profile in the whole non-equilibrium region ($0.5\la r/r_t\la 2$). Turnaround radii are proportional to virialization radii ($r_t\simeq 3.5 r_v$), while turnaround masses are proportional to virialization masses, i.e. $M_t\simeq 1.7 M_v$, where $M_v$ is the mass within $r_v$. Actually, the mass evaluated within the turnaround radius is a more exhaustive evaluation of the total mass of the cluster. These results can be applied to the analysis of observed clusters.
\end{abstract}

\begin{keywords}
cosmology: large scale structure -- galaxies: clusters: general -- galaxies: kinematics and dynamics
\end{keywords}

\section{Introduction}

The gravitational collapse of galaxies towards the centre of clusters is usually described within the framework of the spherical infall model, as the motion of a set of concentric, spherically symmetrical mass shells \citep{GG,Si,Sc}. Actually the spherical infall model is widely accepted in literature, since it describes fairly well the dynamics of the non-equilibrium region of galaxy clusters, defined as the region where the effects of virialization and the crossing of the above-mentioned shells are negligible and some overall infall motion of member galaxies is recognizable. Under the spherical symmetry assumption, the infall motion produces a pattern of caustic surfaces in the galaxy redshift-space distribution (which is obtained representing the line-of-sight velocities of galaxies \emph{versus} their projected position on the sky plane). These caustics envelop all galaxies whose infall motion overwhelms the Hubble flow \citep{K}. Caustics with a characteristical ``trumpet'' shape were actually observed in the redshift-space distribution of clusters \citep{Oal}. Diaferio \& Geller \citep{DG} and Diaferio \citep{D} showed that the caustic amplitude provides a direct measure of the escape velocity of galaxies, and therefore allows to estimate the mass profile of the cluster in the innermost part of the non-equilibrium region, up to the turnaround radius $r_t$ (i.e. the radial distance where the velocity of the infall motion is equal to the Hubble flow velocity.). The caustic technique was applied to the observation of many local clusters \citep{GDK,Ral1,Ral2,Ral3}. These mass estimates are consistent with those based on virial theorem \citep{Giral,BivGir} and weak lensing observations (\citealt{DGR}, and references therein). In fact, up to now the sampled volumes were always restricted within the turnaround radius, and this is due to the definition of caustics surfaces \citep{RG}.

In this paper, we discuss an approach to the issue of mass estimation, which can be applied to larger sampled volumes, well beyond the turnaround radius. We use the radial velocity of galaxies as the key quantity, instead of the escape velocity, as in the caustic technique. According to Silk \citep{Si}, Peebles \citep{P1,P2}, and Gunn \citep{G}, within the spherical symmetry hypothesis, the velocity of the matter infall motion at a certain distance from the centre depends on the encompassed mass. Our purpose is to use this dependence to constrain the value of the overdensity at the turnaround radius. In fact, we see that the turnaround radius is far outside the virialization core of clusters, and is therefore a suitable normalization scale for the cluster mass profile in the non-equilibrium region \citep{VH}. To test our assumptions and to verify the results, we will analys a large galaxy population extracted from a simulated cluster catalogue \citep{Borgal,Bivial}. We will study all clusters both as a whole and one by one. We will prove that the actual turnaround overdensity of clusters is in good agreement with the predictions of the spherical infall model. Moreover, we will show that the normalized mass profiles are generally consistent with a power-law profile, which extends the standard Navarro--Frenk--White profile \citep[hereafter NFW]{NFW1,NFW2,NFW3} to the non-equilibrium region. 

In Section \ref{sec:model} we present the details of our model, concerning the theoretical framework (\ref{sec:theory}) and the simulated data sample (\ref{sec:data}). In Section \ref{sec:results} we discuss the results of our analysis, focusing on the mass estimation at the turnaround radius (\ref{sec:turn}) and in the whole non-equilibrium region, up to 8 virialization radii (\ref{sec:prof}). Finally, in Section \ref{sec:concl} we draw the conclusions of our work. 

\section{The model}\label{sec:model}

\subsection{Theoretical framework}\label{sec:theory}

Consider a galaxy located at a distance $r$ from the centre of a cluster. We call `infall velocity' $v_r$ the peculiar velocity of the galaxy along the radial direction (i.e. towards the cluster centre), assuming that it is positive when directed inwards. The matter overdensity $\delta$ is defined with respect to the background density $\rho_{\itl{bg}}=\Omega_0\rho_{\itl{cr}}$ as follows:
\be
\delta(r)=\f{3}{4\pi r^3}\f{M(r)}{\rho_{\itl{bg}}}-1=\f{3}{\Omega_0\rho_{\itl{cr}} r^3}\left(\int_0^r{\rho r'^2dr'}\right)-1,
\ee 
where $\Omega_0$ is the cosmological matter density parameter and $\rho_{\itl{cr}}$ is the critical density; all the quantities are considered at the present day. According to the hypotheses of the spherical infall model, the ratio between the infall velocity and the Hubble flow velocity $H_0 r$ (where $H_0$ is the Hubble parameter) can be written unambiguously as a function $F$ of both $\Omega_0$ and $\delta$ \citep{Si,P1,G,P2}:
\be\label{eq:F}
\f{v_r}{H_0r}=F(\Omega_0,\delta).
\ee
Several definition of $F$ were proposed in literature. Reg\H os and Geller \citep{RG} demonstrated that quite for all purposes $F$ may be factored into a polynomial $P$ of the mere overdensity $\delta$: $F(\Omega_0,\delta)\simeq\Omega_0^{0.6}P(\delta)$. Lightman \& Schechter \citep{LS} described a simple approach to compute the high-order polynomial terms. However, this formulation is devised to fit the spherical infall in the very-low-overdensity region ($\delta\la 2$). A better agreement in the whole non-equilibrium region ($\delta\la 30$) is obtained with non-polynomial approximations \citep[see Section \ref{sec:turn}]{Y,VD}. Lahav et al. \citep{Lal} took into account the possibility of a non-zero cosmological constant parameter $\Lambda_0$ at the present day, and obtained a corrective term accounting for a 3-per-cent discrepancy with the previous results. Due to the small size of this correction, we will not consider here the effect of $\Lambda_0$, and will assume hereafter that $F$ is approximately factorable into a cosmological term $\Omega_0^{0.6}$ and a generical function $f$ of $\delta$:
\be\label{eq:f}
\f{v_r}{H_0 r}\simeq\Omega_0^{0.6}f(\delta).
\ee

Equation (\ref{eq:F}) and equation (\ref{eq:f}) were commonly used to evaluate $\Omega_0$ from observations of local clusters \citep{RG,LLB,Lal}. Vice versa, since we are handling a simulation and therefore we do know the cosmology, we can reverse this approach and use the equations to compute $\delta$ as a function of $v_r/H_0 r$. In principle, in a purely spherically-symmetric scenario, this would constrain the whole overdensity profile along $r$, because there would be a one-to-one dependence between $r$ and $v_r/H_0 r$. But this assumption is actually too restrictive to describe the overall infall motion of galaxies, even in the non-equilibrium region. In fact, the presence of small-scale substructure is proved to locally affect the galaxy velocity, introducing a sort of random ``kinematical'' noise which blurs the infall velocity profile \citep{DG}. Nevertheless, we will prove that this profile is regular enough to make possible the estimate of the turnaround radius $r_t$, which is defined by the condition $v_r/H_0 r_t=1$. If so, equation (\ref{eq:f}) can be used to implicitly define the turnaround overdensity $\delta_t\equiv\delta(r_t)$ as a function of $\Omega_0$:
\be\label{eq:deltat_imp}
f(\delta_t)\simeq\Omega_0^{-0.6}.
\ee

Since we are using the cosmological N-body simulation described in Section \ref{sec:data}, we have analyzed the dependence on $\Omega_0$ by means of analytical calculations, which assume spherically symmetric infall, and a spatially flat Universe. This model provides an analytical definition of $r_t$ and $\delta_t$ for a cluster (\citealt{ECF,Lal}; see App. \ref{app:spherical_flat}):
\be\label{eq:rt_spherical}
r_{t}=\f{2-\eta_v/2}{1-\eta_v/2}\f{\kappa_t^{1/2}\cos\theta_t}{\kappa_v^{1/2}\cos\theta_v}r_v,
\ee
\be\label{eq:deltat_spherical}
1+\delta_{t}=\left[\f{3(1-\Omega_0)}{4\kappa_t\Omega_0\left(\cos\theta_t\right)^2}\right]^{3/2}.
\ee
In these equations, $r_v$ is the virialization radius of the cluster, while the constants $\kappa_v$ and $\kappa_t$ parametrize the amplitude of two density perturbations which are respectively collapsing and turning around at the present day. The angle $\theta_v$ and the parameter $\eta_v$ are both related to $\kappa_v$, while the angle $\theta_t$ is related to $\kappa_t$ (see App. \ref{app:spherical_flat} for further details). In principle, all parameters in equation (\ref{eq:rt_spherical}) and equation (\ref{eq:deltat_spherical}) depend on the adopted cosmology, and in particular on the value of $\Omega_0$. In the range $0.2\le\Omega_0\le 0.4$, the dependence on $\Omega_0$ is approximately factorable as follows (with a one-per-cent accuracy):
\be\label{eq:rt_spherical_simple}
r_{t}\simeq\Omega_0^{1/4}\tilde r_{t},
\ee
\be\label{eq:deltat_spherical_simple}
1+\delta_{t}\simeq\Omega_0^{-3/4}(1+\tilde\delta_{t}),
\ee
where $\tilde r_t$ and $\tilde\delta_t$ are independent from cosmology. According to the latest WMAP observations \citep{Spal}, $\Omega_0=0.27\pm 0.04$. Substituting this value into equation (\ref{eq:rt_spherical}) and equation (\ref{eq:deltat_spherical}), we get $r_t=(3.02\pm 0.02) r_v$ and $\delta_t=12.2_{-1.3}^{+1.6}$, corresponding to $r_t=(0.72\pm 0.03)\tilde r_t$ and $\delta_t=(2.7\pm 0.3)\tilde\delta_t$ from equation (\ref{eq:rt_spherical_simple}) and equation (\ref{eq:deltat_spherical_simple}), where $\tilde r_t=4.2$ and $\tilde\delta_t=4.6$. As one can see, the uncertainty on $r_t$ and $\delta_t$ due to  dependence on cosmology is very small and can be neglected for our purposes (as it will be shown in Section \ref{sec:turn}). For these reason, we will adopt hereinafter the concordance value $\Omega_0=0.3$ if not otherwise specified. 

The turnaround radius $r_t$ will be adopted as a normalization scale useful to describe the outskirts of clusters; in this way, it replaces the virialization radius, which is usually adopted in the cluster core. Within this framework, we will demonstrate that in the non-equilibrium region the overdensity profile $\delta_{\rmn{NE}}(r)$ and the mass profile $M_{\rmn{NE}}(r)$ of a given cluster are generally consistent with a single profile, if they are normalized to the turnaround scale. Therefore, we can write:
\be\label{eq:delta_NE}
\delta_{\rmn{NE}}(r)=(1+\delta_t)\left(\f{r}{r_t}\right)^{-3}g_{\rmn{NE}}(r)-1,
\ee
\be\label{eq:M_NE}
M_{\rmn{NE}}(r)=M_t g_{\rmn{NE}}(r)=\f{4}{3}\pi r_t^3\Omega_0\rho_{\itl{cr}}(1+\delta_t)g_{\rmn{NE}}(r),
\ee
where $g_{\rmn{NE}}$ is a function to be defined. Equation (\ref{eq:delta_NE}) and equation (\ref{eq:M_NE}) formally correspond to the equations of the Navarro-Frenk-White profile \citep[hereafter NFW]{NFW1,NFW2,NFW3}. The difference between our profile and the NFW one lies in the choice of the normalization scale (the turnaround radius $r_t$ and the turnaround overdensity $\delta_t$ substitute the virialization radius $r_v$ and the virialization overdensity $\delta_v$), and lies also in the definition of $g_{\rmn{NE}}$, which we will see to be different from the corresponding NFW function $g_{\rmn{NFW}}(r)=\ln(1+c_v r/r_v)-c_v r/(c_v r+r_v)$, where $c_v$ is the cluster concentration parameter \citep{Bal,LM}. 
 
\subsection{The simulated catalogue}\label{sec:data}

\begin{figure}
\centering
\includegraphics[scale=0.4]{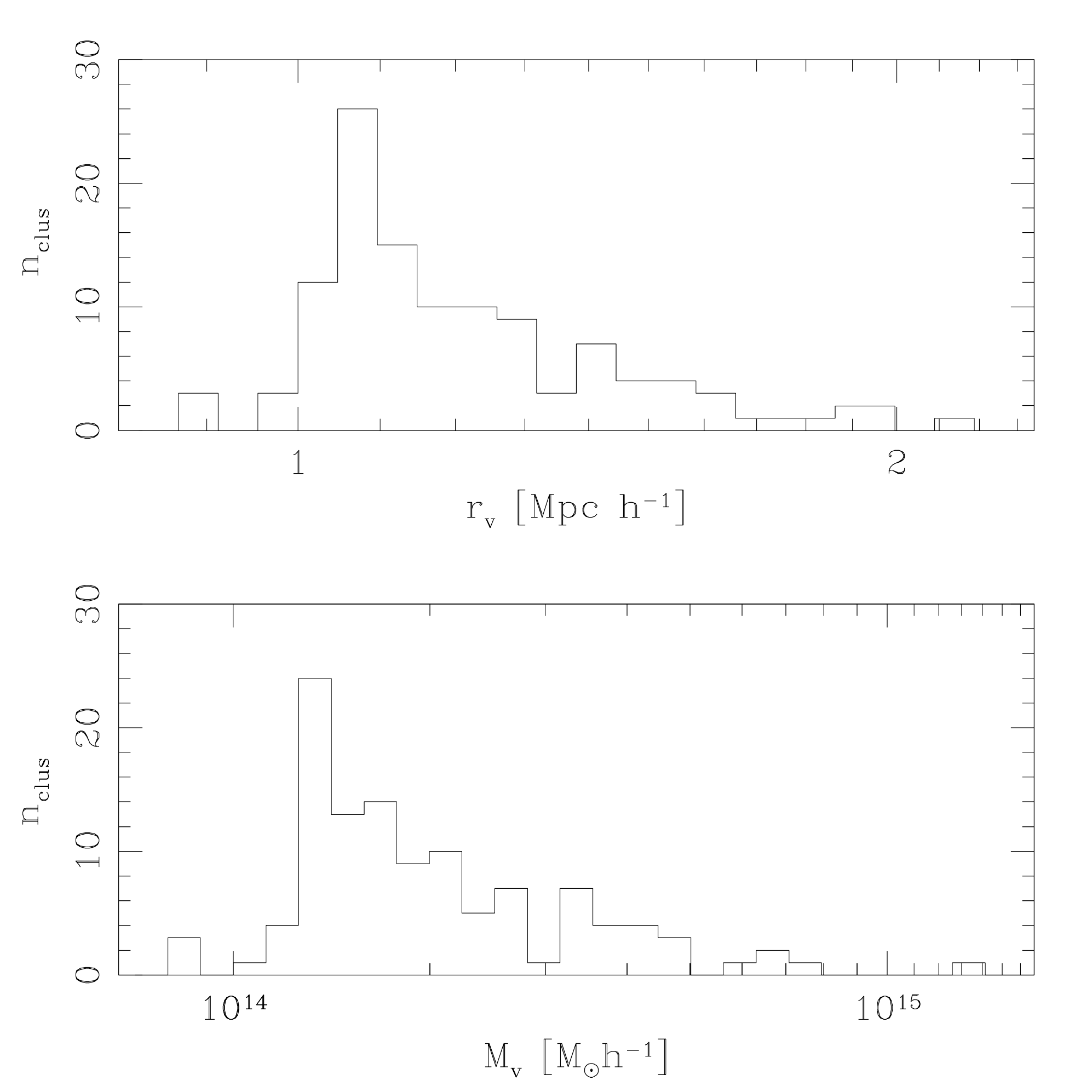}
\caption{Characteristics of the sample. First plot: frequency distribution of the virialization radii $r_v$; second plot: frequency distribution of the virialization masses $M_v$. $n_{clus}$ is the number of cluster per frequency bin.}
\label{fig:fdistr_char}
\end{figure}

The model was tested on a catalogue of 114 simulated clusters, with an overall population of 9631 galaxies. The clusters and the galaxies were extracted by Biviano et al. \citep{Bivial} from a large cosmological hydrodynamical simulation performed by Borgani et al. \citep{Borgal}. We refer to these papers for a detailed description of the data sample. We just remark what follows:
\begin{enumerate}
\p The simulation was run with the tree-\textsc{sph} \textsc{gadget}-2 code \citep{SYW,SH}, adopting a $\Lambda$-CDM cosmology ($\Omega_0=0.3$, $\Omega_\Lambda=1-\Omega_0$, $\Omega_{\itl{bar}}=0.019h^{-2}$, $h=0.7$ and $\sigma_8=0.8$). It traced the evolution of $480^3$ dark matter (DM) particles and $480^3$ gas particles (partly converted into stellar particles during the run) within a box of volume $(192h^{-1})^3$ Mpc$^{3}$. 
\p The clusters were identified at $z=0$ with a standard Friends-of-Friends (FoF) algorithm, taking into account the DM particles of the simulation. After the identification, a spherical overdensity algorithm was applied to determine the size of the virialization core of each cluster. The virialization overdensity was defined as follows, in agreement with the adopted cosmology \citep{BN}:
\be
(1+\delta_v)\Omega_0=18\pi^2+82(\Omega_0-1)-39(\Omega_0-1)^2\simeq 101.
\ee
\p The galaxies were identified with the publicly available algorithm \textsc{skid} \citep{S}; in this case, only the stellar component was taken into account. 
\end{enumerate}

According to the definition of $\delta_v$, we will define the virialization radius and the virialization mass of each cluster as $r_v\equiv r_{101}$ and $M_v\equiv M_{101}$, respectively. The extracted clusters are very different in size, with $r_v$ ranging from $0.88h^{-1}$ Mpc to $2.23h^{-1}$ Mpc and $M_v$ ranging from $7.95\times 10^{13}h^{-1}M_\odot$ to $1.30\times 10^{15}h^{-1}M_\odot$. The frequency distribution of $r_v$ and $M_v$ among the sample is shown in Fig. \ref{fig:fdistr_char}. Since many authors prefer to use $r_{200}$ and $M_{200}$ instead of $r_v$ and $M_v$, we provide the average ratios $r_v/r_{200}$ and $M_v/M_{200}$ computed on the entire cluster catalogue: 
\be
\f{r_v}{r_{200}}=1.36\pm 0.04,
\ee
\be
\f{M_v}{M_{200}}=1.26\pm 0.11.
\ee

The number of member galaxies is very different in different cluster, ranging from $17$ to $403$. To make different objects comparable, we sliced all clusters into a set of concentric shells, using the virialization radius as the scale reference. The shells were defined so as to cover the whole extent from the virialization core to the far outskirts of clusters (i.e. from $0.1r_v$ to $8r_v$). We adopted a logarithmical spacing in order to fit the decreasing galaxy number density along the radial coordinate. We defined the outer radius $r_j$ of each shell $j$ as follows:
\be
r_j=r_v10^{(j/50)-1},\qquad j=1,\ldots,91.
\ee
(The choice of 91 shells is technical, induced by the quality of our data.) The same spacing was used to reconstruct the matter distribution along the radial coordinate in the clusters. 

The integrated overdensity and mass profile were computed taking into account all the particles in the simulation (i.e., DM, gas, and stellar particles), while the infall velocity was extracted from the member galaxies alone. We adopted this approach to better investigate the possibility of reconstructing the cluster mass distribution using only the dynamical properties of the member galaxies, which at least in principle can be directly inferred from the observations. Within this approach, $\delta_j$ and $M_j$ are defined respectively as the overdensity and the mass of all the particles enclosed within the sphere of radius $r_j$, while $v_{r;j}$ is the mean infall velocity of all the galaxies within the shell $j$. We used the DM particles as a tracer of the cluster dynamics only when computing the 3-d velocity dispersion within the virialization core of clusters (Section \ref{sec:turn}), because in this case the DM component yields a stabler result due to its larger statistical significance.

\section{The results}\label{sec:results}

\subsection{The turnaround radius and the overdensity estimation}\label{sec:turn}

\begin{figure}
\centering
\includegraphics[scale=0.4]{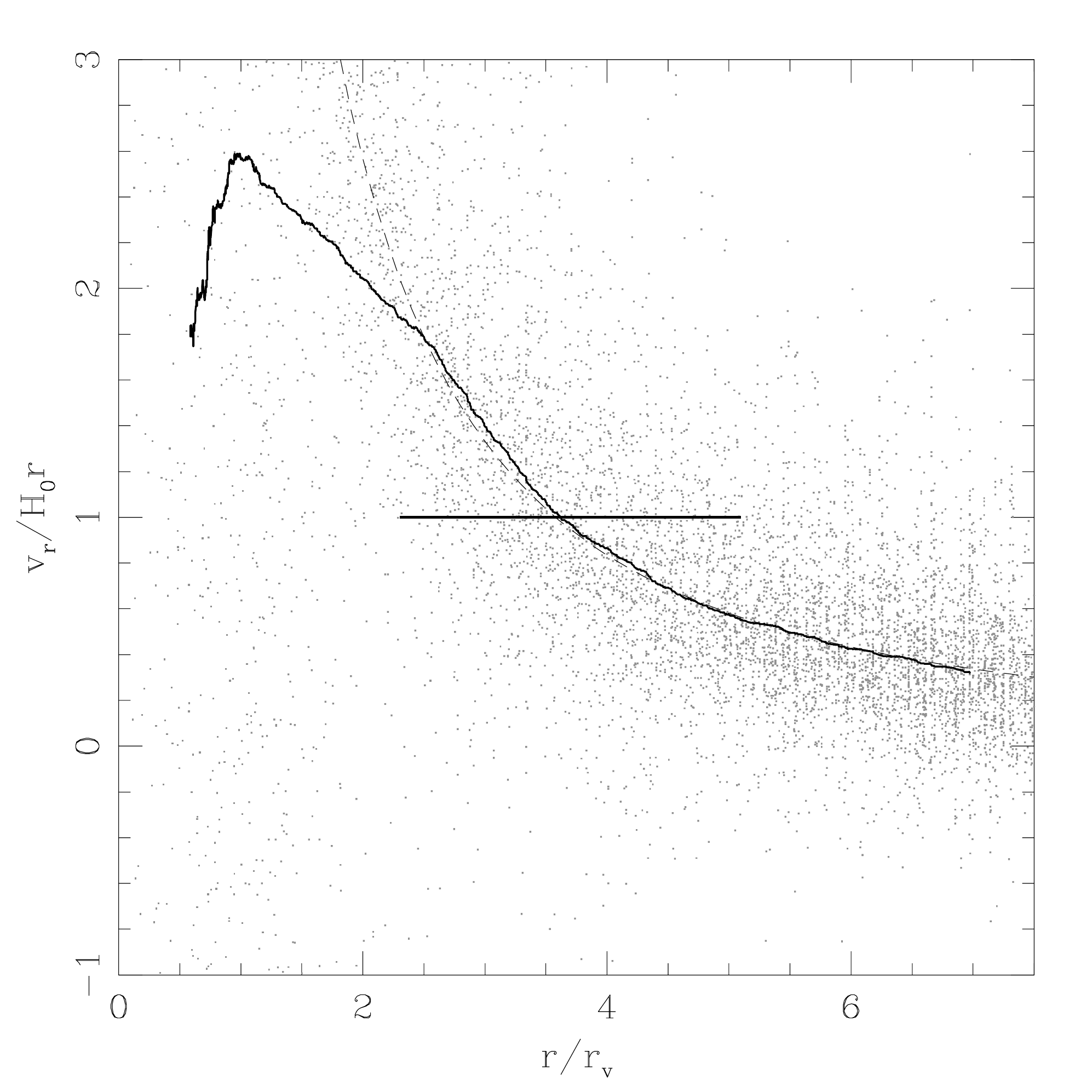}
\caption{Overall infall velocity profile of member galaxies as a function of the normalized radial distance. The distribution of galaxies (points) has been smoothed with a running median (thick solid line) and interpolated with a power law (dashed line; see text). The horizontal bar indicates the turnaround condition $v_r/H_0 r=1$.}
\label{fig:inf_s}
\end{figure}

\begin{figure}
\centering
\includegraphics[scale=0.4]{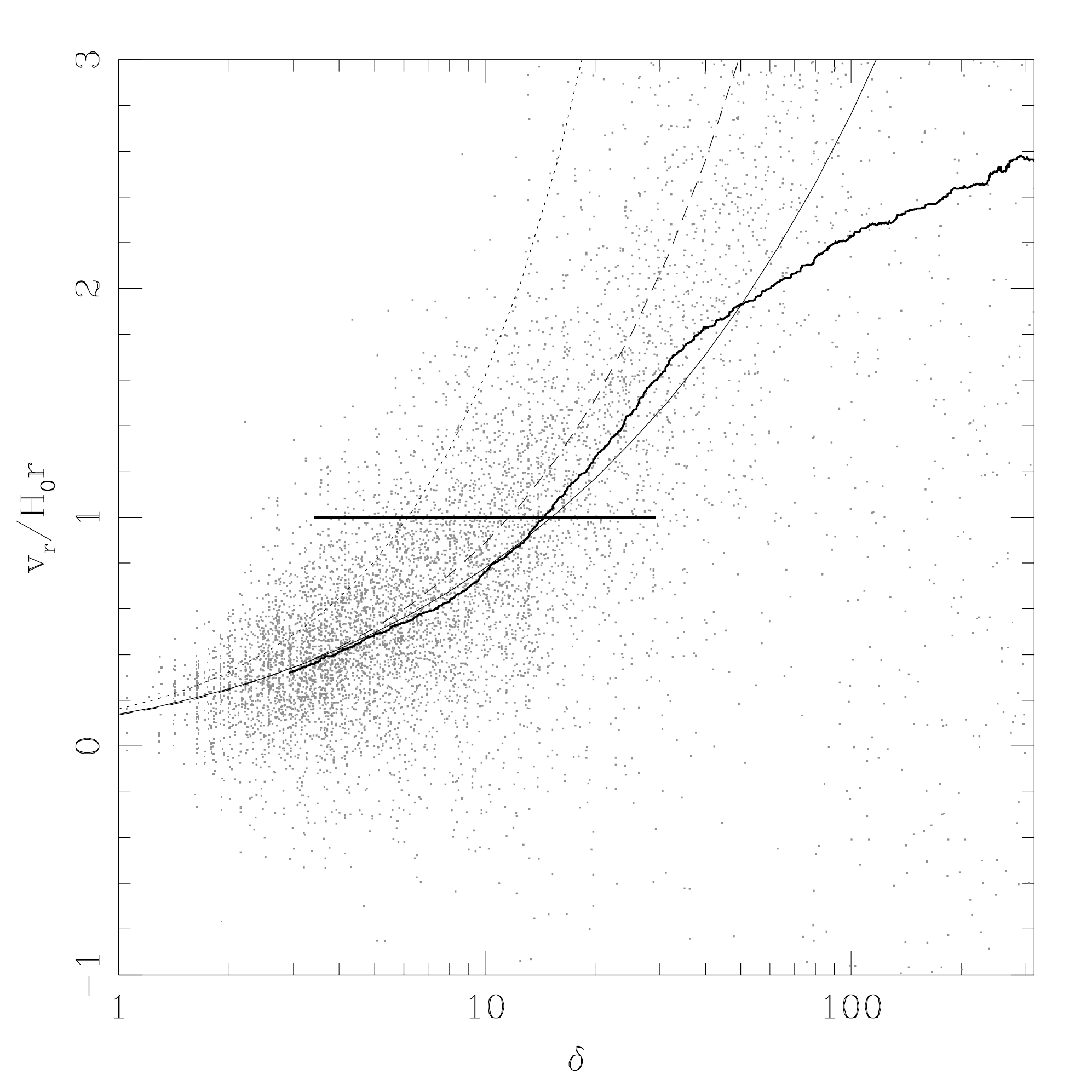}
\caption{Overall infall velocity profile of member galaxies as a function of the overdensity. The distribution of galaxies (points) has been smoothed with a running median (thick solid line) and then interpolated with $f_{\itl{lin}}$ (dotted line), $f_{\rmn{Y}}$ (dashed line), and $f_{\rmn{M}}$ (narrow solid line). The horizontal bar indicates the turnaround condition $v_r/H_0 r=1$.}
\label{fig:inf_delta}
\end{figure}

Fig. \ref{fig:inf_s} and Fig. \ref{fig:inf_delta} represent the sample distribution of the normalized infall velocity $v_r/H_0 r$ of member galaxies as a function of the normalized radial distance $r/r_v$ and the overdensity $\delta$, respectively. We superimposed all clusters into a single synthetic object, in order to increase the statistical significance, as suggested by \citet{VH}. Each point corresponds to a single galaxy, while the thick solid line is the running median (RM) of the distribution. Although the large variance, we can well recognize a common profile in the intervals $r\ga 2.2r_v$ and $\delta\la 35$. This trend indicates the existence of an overall and well-defined galaxy infall motion in the non-equilibrium region of clusters.

In both Fig. \ref{fig:inf_s} and Fig. \ref{fig:inf_delta}, we use the RM as a reference profile to describe the overall dynamics in the non-equilibrium region, in order to find the values of the turnaround radius and the turnaround overdensity. The radius $r_t$ is computed by interpolating the RM profile of Fig. \ref{fig:inf_s} with a power law (dashed line). A linear-fitting algorithm applied to the bilogarithmic distribution gives:
\be\label{eq:PL}
\log_{10}\left(\left.\f{v_r}{H_0 r}\right|_{\rmn{RM}}\right)=\alpha_{v_r}+\beta_{v_r}\left(\f{r}{r_v}\right),
\ee
where $\alpha_{v_r}=0.96\pm 0.03$ and $\beta_{v_r}=1.72\pm 0.02$ (1-$\sigma$ uncertainties). Equation (\ref{eq:PL}) is in good agreement with the RM profile for $r\ga 2.2r_v$, and can be used to compute the median turnaround radius $r_{t,\rmn{RM}}$, defined by the condition $\left.v_r/H_0 r_t\right|_{\rmn{RM}}=1$ (corresponding to the horizontal bar in Fig. \ref{fig:inf_s}). We obtain:
\be\label{eq:st_RM}
r_{t,\rmn{RM}}=(3.61\pm 0.02)r_v.
\ee 
In this equation, the 1-$\sigma$ uncertainty is due to the interpolation algorithm and does not take into account the variance among the clusters in the sample, which will be considered later on. 

To determine $\delta_t$, we compare the RM profile in Fig. \ref{fig:inf_delta} with the plots of three different expressions of the function $f$ defined in equation (\ref{eq:f}), namely the linear approximation $f_{\itl{lin}}$ \citep[dotted line]{P1,G,P2}, the Yahil approximation $f_{\rmn{Y}}$ \citep[dashed line]{Y}, and the Meiksin approximation $f_{\rmn{M}}$ \citep[narrow solid line]{VD}:
\be\label{eq:lin}
f_{\itl{lin}}(\delta)\equiv\f{1}{3}\delta,
\ee
\be\label{eq:Y}
f_{\rmn{Y}}(\delta)\equiv\f{1}{3}\delta(1+\delta)^{-1/4},
\ee
\be\label{eq:M}
f_{\rmn{M}}(\delta)\equiv\f{1}{3}\delta\left(1+\f{1}{3}\delta\right)^{-1/2}.
\ee
According to equation (\ref{eq:deltat_imp}), these expressions provide as many implicit definition of the turnaround overdensity, corresponding to the points of intersection, in Fig. \ref{fig:inf_delta}, of the three curves with the horizontal bar $v_r/H_0 r=1$. As one can see from Fig. \ref{fig:inf_delta}, the linear approximation poorly describes the infall motion in the non-equilibrium region, since it departs from the data distribution even in the low overdensity region. Conversely, the non-linear functions $f_{\rmn{Y}}$ and $f_{\rmn{M}}$ are close to RM up to the turnaround region. Equation (\ref{eq:lin}) and equation (\ref{eq:M}) can be inverted by trivial algebraic computation, while equation (\ref{eq:Y}) requires an \emph{ad hoc} treatment (see App. \ref{app:Y_inv}). The results obtained in the three cases are, respectively,
\be\label{eq:deltat_lin}
\delta_{t,\itl{lin}}=3\Omega_0^{-0.6}\simeq 6,
\ee
\be\label{eq:deltat_Y}
\delta_{t,\rmn{Y}}\simeq \f{66}{17}11^{1/4}\Omega_0^{-0.6}\f{v_r}{H_0 r}-\f{50}{17}\simeq 11,
\ee
\be\label{eq:deltat_M}
\delta_{t,\rmn{M}}=\f{3}{2}\Omega_0^{-1.2}\left[1+\sqrt{1+4\Omega_0^{1.2}}\right]\simeq 15.
\ee
It is worth noticing that the Yahil approximation agrees with the prediction of the spherical infall model (see Section \ref{sec:theory}). Conversely, the Meiksin approximation shows the best agreement with the simulated data, being very close to RM profile in the interval $\delta\la 40$. This is an evidence of the disagreement between the purely-spherical description and the actual galaxy dynamics in the non-equilibrium region of galaxy cluster.

\begin{figure}
\centering
\includegraphics[scale=0.4]{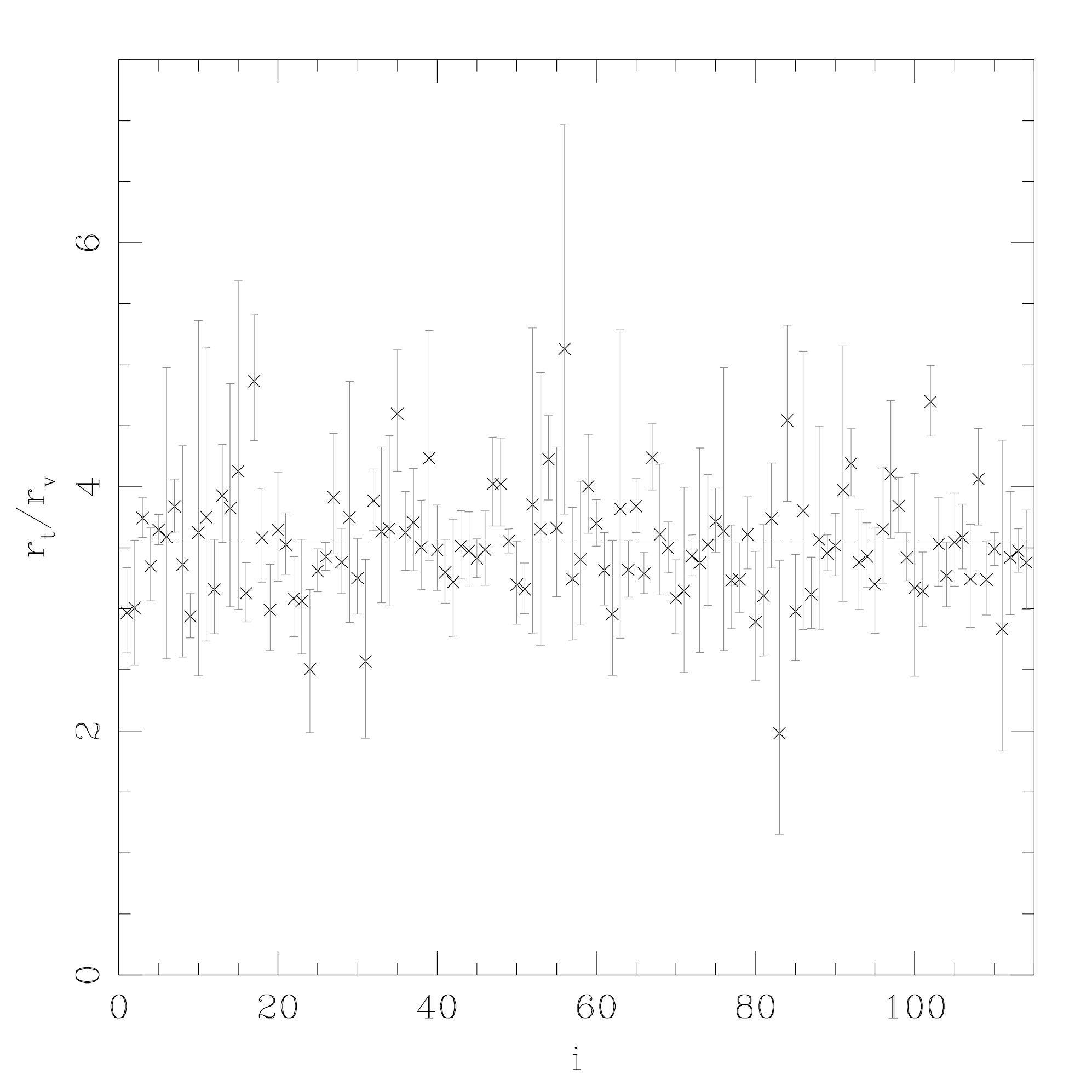}
\caption{Turnaround radius estimation for all clusters in the data sample. The individual value of $r_{t;i}$ extracted from the DM distribution of galaxies (crosses and error bars) is compared with $r_{t,\itl{RM}}$ (dashed line). The error bars (1 $\sigma$) correspond to the uncertainty associated to the fit of the infall velocity profile. $i$ is the index number of each cluster.}
\label{fig:st}
\end{figure}

\begin{figure}
\centering
\includegraphics[scale=0.4]{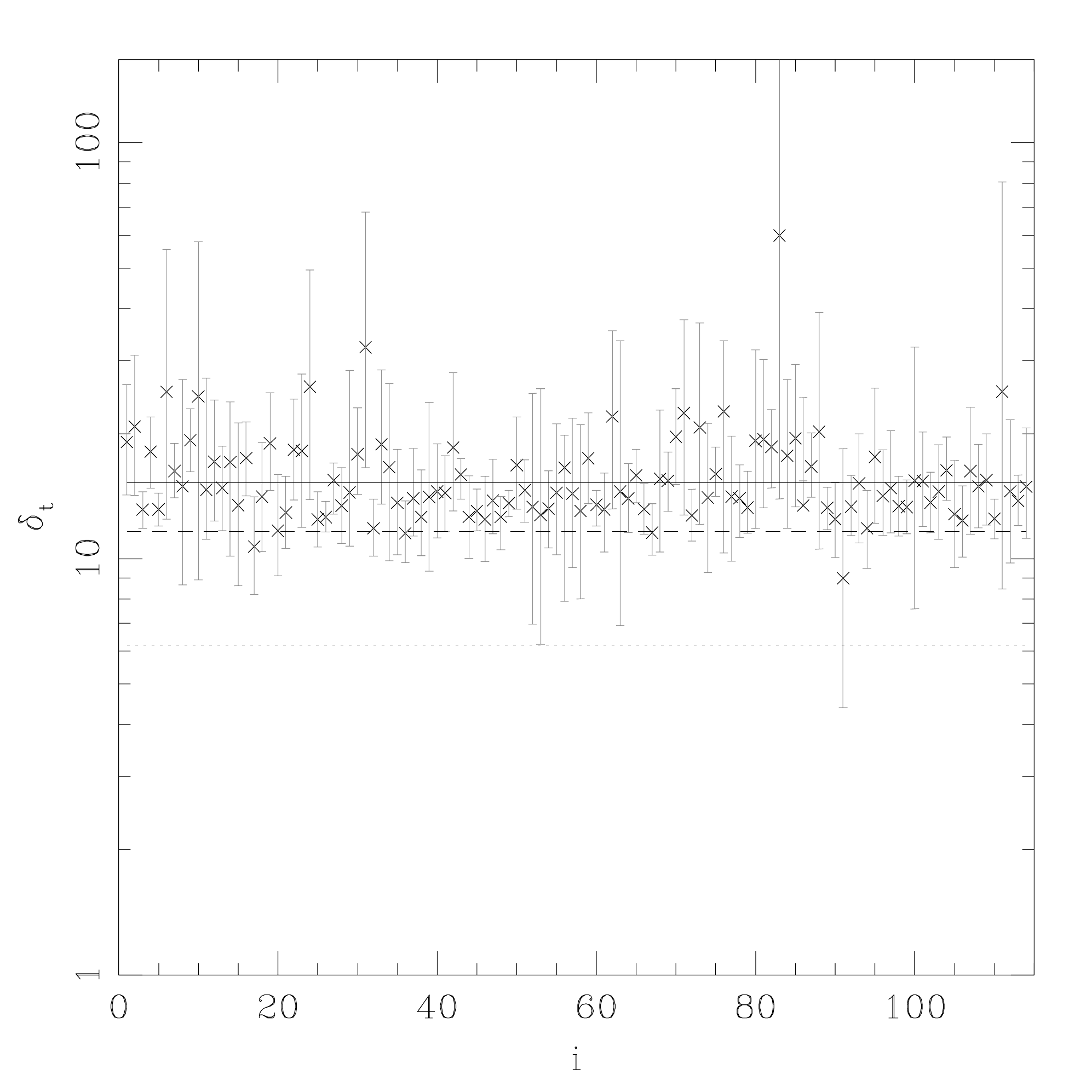}
\caption{Turnaround overdensity estimation for all clusters in the data sample. The value of $\delta_{t;i}$ extracted from DM distribution (crosses and error bars) is compared with $\delta_{t,\itl{lin}}$ (dotted line), $\delta_{t,\rmn{Y}}$ (dashed line), and $\delta_{t,\rmn{M}}$ (solid line). The error bars (1 $\sigma$) correspond to the uncertainty associated to the fit of the infall velocity profile. $i$ is the index number of each cluster.}
\label{fig:deltat}
\end{figure}

We also estimated the value of $r_t$ and $\delta_t$ for single clusters separately. In this case, we traced the infall velocity profile using the DM particles instead of the galaxies, since the galaxy distribution is typically noisier (because in a single cluster the number of galaxies is much smaller than the number of DM particles traced by the simulation). Fig. \ref{fig:st} and Fig. \ref{fig:deltat} represents the individual values $r_{t;i}$ and $\delta_{t;i}$, respectively, as a function of the index number $i$ of our cluster catalogue (crosses and error bars). The error bars (1 $\sigma$) correspond to the uncertainty associated to the fit of the infall velocity profile. Most of values lie in a quite narrow band. In particular, the turnaround radius is (in almost all cases) not only larger but considerably larger than the virialization radius, so it lies in the infall region, and therefore it can be adopted as a suitable normalization scale for the non-equilibrium region. Averaging the individual values over the whole sample we obtain:
\be\label{eq:st}
\log_{10}\left(\f{\overline{r}_t}{r_v}\right)=0.54\pm 0.05,
\ee
\be\label{eq:deltat}
\log_{10}\left(1+\overline{\delta}_t\right)=1.2\pm 0.1.
\ee
corresponding to $\overline{r}_t=(3.5\pm 0.4) r_v$ and $\overline{\delta}_t=15_{-3}^{+4}$. The uncertainty on these estimates is much larger than the uncertainty related to the value of $\Omega_0$ (see Section \ref{sec:theory}); for this reason the latter has been neglected, and the concordance value $\Omega_0=0.3$ has been used throughout. There is a highly significant correlation between $\log_{10}\overline{r}_t$ and $\log_{10}(1+\overline{\delta}_t)$, as indicated by the Pearson's correlation coefficient $r_{\rmn{P}}=-0.61$ (significance $\gg 99\%$). This result is not surprising, since a mutual dependence between $v_r$ and $r$, and between $v_r$ and $\delta$, yields naturally a mutual dependence between $r_t$ and $\delta_t$. Taking into account this correlation, we can estimate the turnaround mass $\overline{M}_t=M_v(\overline{r}_t/r_t)^3(1+\overline{\delta}_t)/(1+\delta_v)$ with the correct 1-$\sigma$ uncertainty, as follows:
\be\label{eq:Mt}
\log_{10}\left(\f{\overline{M}_t}{M_v}\right)=0.24\pm 0.01,
\ee 
corresponding to $\overline{M}_t=(1.74\pm 0.04)M_v$. Equation (\ref{eq:Mt}) confirms the estimate of  \citet{RD}, which measured the average ratio $M_t/M_{200}$ by analysing a sample of observed clusters.

The value of $\overline{r}_t$ from equation (\ref{eq:st}) is in agreement with the value of $r_{t,\rmn{RM}}$ from equation (\ref{eq:st_RM}) (dashed line in Fig. \ref{fig:st}), and the 1-$\sigma$ uncertainty on the turnaround radius now takes into account the variance among the clusters. On the other hand, equation (\ref{eq:deltat}) rules out the linear approximation result $\delta_{t,\itl{lin}}$ (dotted line in Fig. \ref{fig:deltat}), and confirms the non-linear estimates $\delta_{t,\rmn{Y}}$ and $\delta_{t,\rmn{M}}$ (dashed and solid line in Fig. \ref{fig:deltat}, respectively). The Meiksin approximation provides again the best estimate; therefore, it will be adopted hereafter as the best expression for $f$. 

The values in equation (\ref{eq:st}) and (\ref{eq:deltat}) are not different within the uncertainties from the corresponding predictions of the spherical infall model. However, the marginal evidence (within 1 $\sigma$) of larger value of $\overline{r}_t$ could be ascribed to a collapse which is not perfectly spherical. In fact, Hoffman (1986) found that a shear in the velocity field can induce higher infall velocities, which lead to larger turnaround radii.

\begin{figure}
\centering
\includegraphics[scale=0.4]{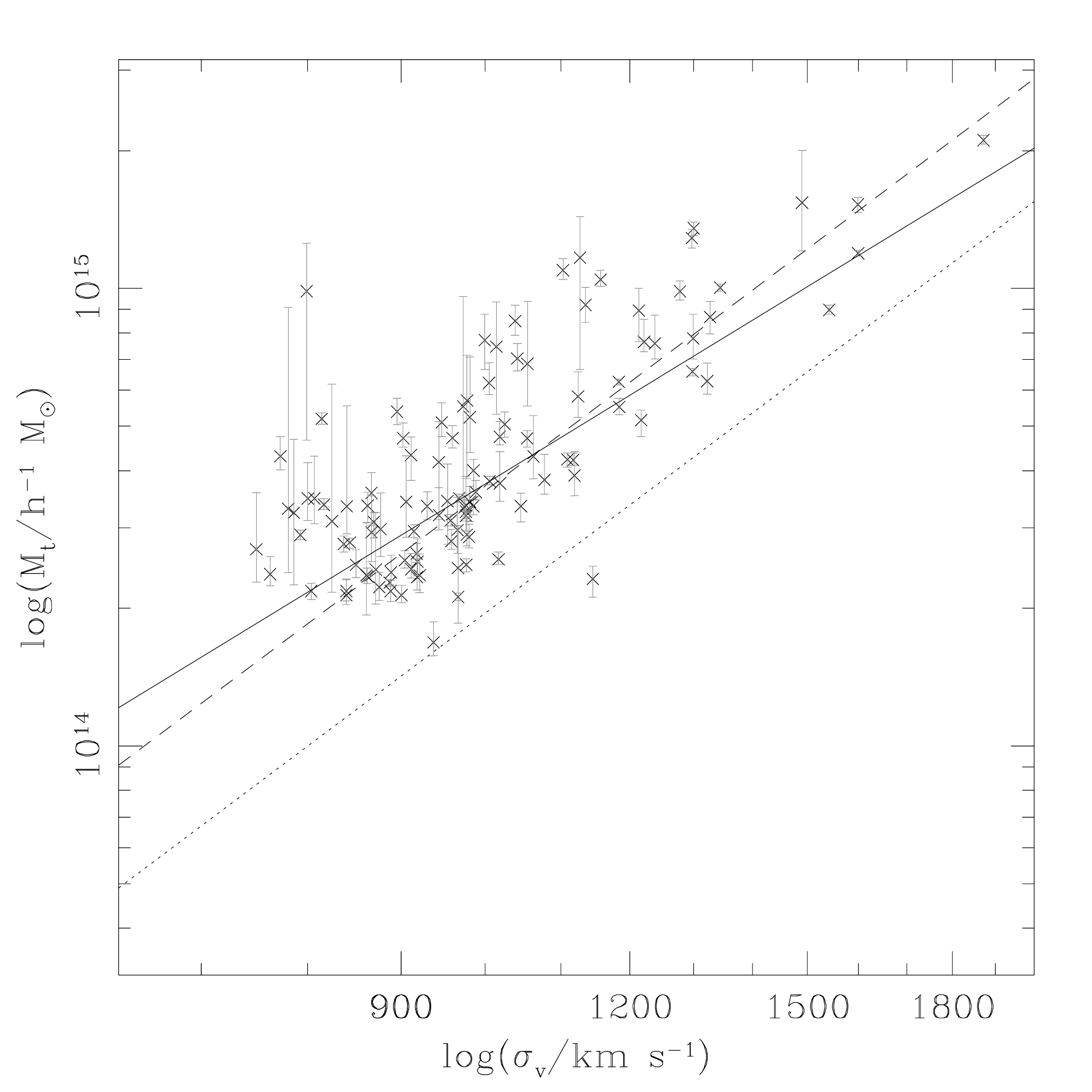}
\caption{Dependence of the estimated turnaround mass on the 3-d DM velocity dispersion within the virialization core. The crosses represent the expected values $(\sigma_{v,i},M_{t;i})$, while the error bars represent the $1$-$\sigma$ uncertainty on the estimation of $M_{t;i}$, where $i$ is the index number of our cluster catalogue. These values are compared with the best-fitting power law (solid line) and the best-fitting cubic relation (dashed line). The cubic relation which best fits the corresponding distribution of virialization masses $M_v$ (Biviano et al. 2006) is represented by the dotted line.}
\label{fig:sigma_Mt}
\end{figure}

Biviano et al. \citep{Bivial} show that the virialization mass $M_v$ depends on the 3-dimensional velocity dispersion of the DM component within the virialization radius, $\sigma_{v,\rmn{DM}}$. Since the ratio $M_t/M_v$ is quite constant in our catalogue, we expect to find a relation between the turnaround mass and $\sigma_{v,\rmn{DM}}$. We therefore compare the estimated values of cluster turnaround masses $M_{t;i}$ with the values of $\sigma_{v,\rmn{DM}}$ extracted from the simulated catalogue. The dependence of cluster turnaround masses on the velocity dispersion is shown in Fig. \ref{fig:sigma_Mt}; the solid line is obtained applying a linear-fitting algorithm to the bilogarithmic distribution, which gives
\be\label{eq:sigma_Mt}
\log_{10}\left(\f{M_t}{10^{14}h^{-1} M_\odot}\right)=\alpha_{M_t}+\beta_{M_t}\log_{10}\left(\f{\sigma_v}{10^3\textrm{km s}^{-1}}\right),
\ee
where $\alpha_{M_t}=0.6\pm 0.1$ and $\beta_{M_t}=2.4\pm 1.3$.
We also computed the best-fitting cubic relation for the same distribution (dashed line):
\be\label{eq:sigma_Mt3}
\log_{10}\left(\f{M_t}{10^{14}h^{-1} M_\odot}\right)=\tilde{\alpha}_{M_t}+3\log_{10}\left(\f{\sigma_v}{10^{3}\textrm{km s}^{-1}}\right),
\ee
where $\tilde{\alpha}_{M_t}=0.73\pm 0.05$. Equation (\ref{eq:sigma_Mt3}) is consistent with equation (\ref{eq:sigma_Mt}) within the uncertainties. The cubic relation is favoured by Biviano et al. \citep{Bivial} to describe the $M_v$-$\sigma_v$ dependence (since $M_v\sim\sigma_v^2 r_v$ and $r_v\sim\sigma_v$), and therefore it is expected to work also to describe the $M_t$-$\sigma_v$ dependence. 
Our value of $\tilde{\alpha}_{M_t}$ is consistent with the corresponding value by Biviano et al. \citep{Bivial}. We point out that in principle it is possible to use equation (\ref{eq:sigma_Mt}) or equation (\ref{eq:sigma_Mt3}) to obtain a mass estimate entirely based on the 3-dimensional velocity dispersion of the DM particles. One could also use galaxies instead of dark matter, provided that the galaxy velocity dispersion is an unbiased estimator of $\sigma_{v,\rmn{DM}}$. This point is still debated in the literature, see, e.g., \citet{Bivial} and references therein. However, in their analysis, \citet{Bivial} found that the bias is negligible when all galaxies (not only early-type galaxies) are considered.

\subsection{Overdensity and mass profile estimation}\label{sec:prof}

\begin{figure}
\centering
\includegraphics[scale=0.4]{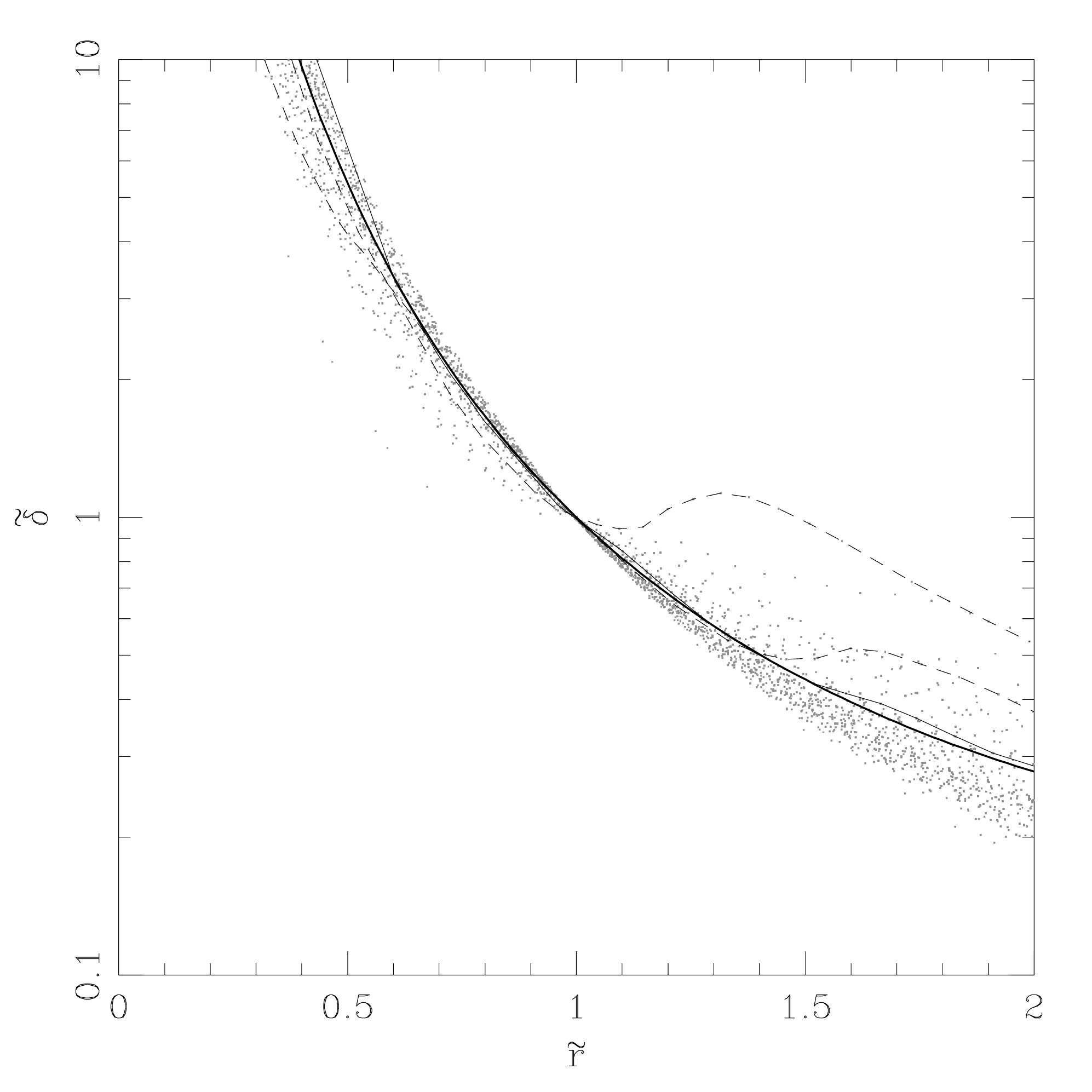}
\includegraphics[scale=0.4]{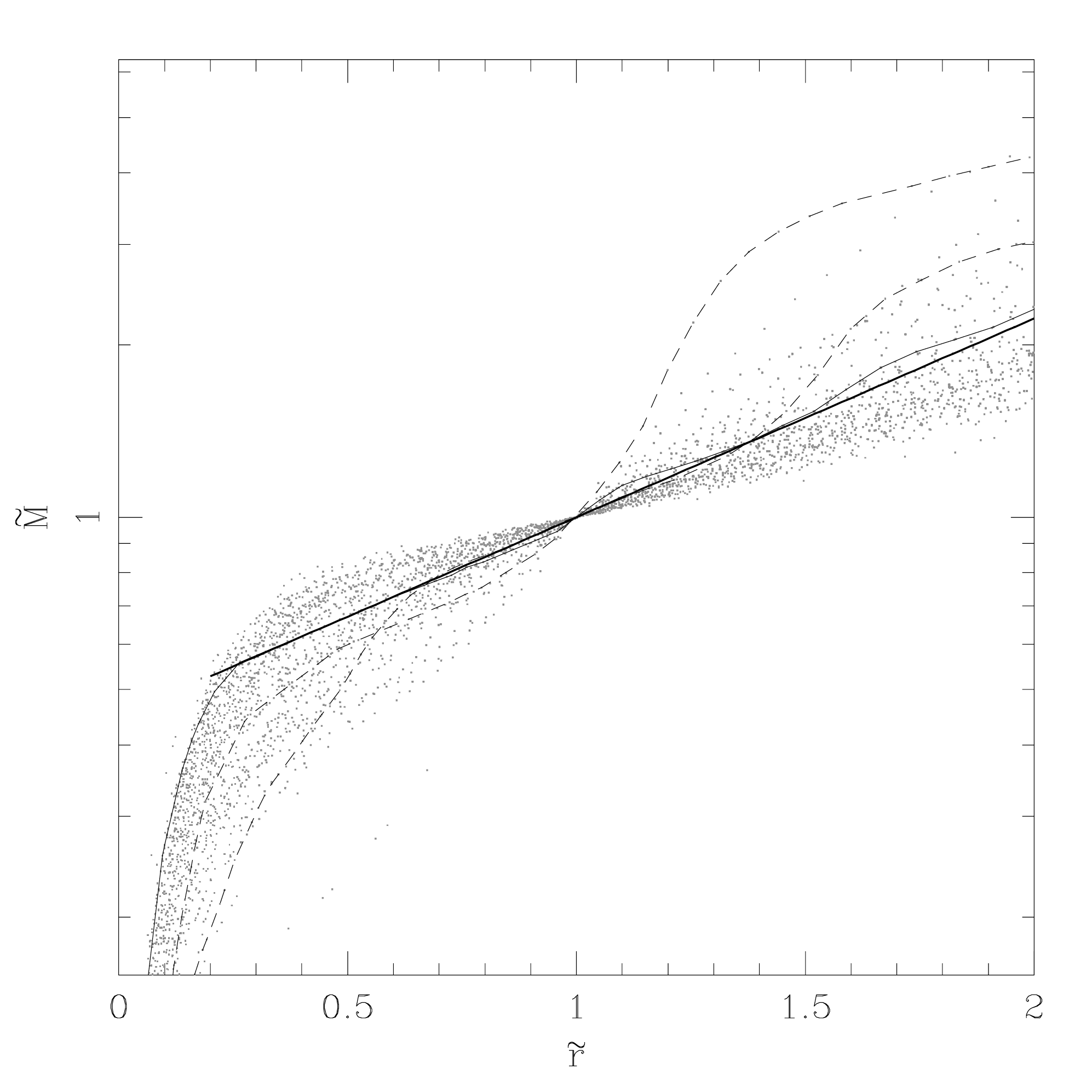}
\caption{Radial profile of the normalized overdensity $\tilde\delta^{(1)}$ and of the normalized mass $\tilde M^{(1)}$ for the whole sample. The profiles were reconstructed through the distribution of member galaxies (points) and subsequently interpolated with $\tilde\delta_{\rmn{NE}}$ and $\tilde M_{\rmn{NE}}$, respectively (solid lines); see text and equations (\ref{eq:norm_i_delta}) and (\ref{eq:norm_i_M}). In both plots, we highlighted a regular profile (narrow solid line), and two irregular profiles (dashed line).}
\label{fig:prof_M}
\end{figure}

Once $r_t$ and $\delta_t$ are known, we use them to normalize the overdensity profile and the mass profile of the clusters. The normalized profiles $\tilde\delta$ and $\tilde M$ are obtained in two different ways ($i$ is the label of the cluster and $j$ is the label of the cluster shell considered; see Section \ref{sec:turn}):
\begin{enumerate}
\renewcommand{\theenumi}{(\arabic{enumi})}
\p Using the individual values of turnaround radius and turnaround overdensity, computed for each cluster separately:
\be\label{eq:norm_i}
\tilde r_{i,j}^{(1)}=\f{r_{i,j}}{r_{t;i}},\qquad\tilde\delta_{i,j}^{(1)}=\f{1+\delta_{i,j}}{1+\delta_{t;i}},\qquad\tilde M_{i,j}^{(1)}=\f{M_{i,j}}{M_{t;i}},
\ee
where $M_{t;i}\equiv M_v(r_{t;i}/r_v)^3(1+\delta_{t;i})/(1+\delta_v)$;
\p Using the mean values of turnaround radius and turnaround overdensity, obtained from the data of the whole cluster sample:
\be\label{eq:norm_ii}
\tilde r_{i,j}^{(2)}=\f{r_{i,j}}{r_\rmn{t,RM}},\qquad\tilde\delta_{i,j}^{(2)}=\f{1+\delta_{i,j}}{1+\delta_{t,\rmn{M}}},\qquad\tilde M_{i,j}^{(2)}=\f{M_{i,j}}{M_{t,\rmn{M}}},
\ee
where $M_{t,\rmn{M}}\equiv M_v(r_{t,\rmn{RM}}/r_v)^3(1+\delta_{t,\rmn{M}})/(1+\delta_v)$.
\end{enumerate}
These two normalization criteria are useful to test the reliability of our model when applied to a sample of imulated  clusters, as we will discuss later.

\begin{table}
\caption{Value of the parameters in equation (\ref{eq:delta_ti}) for different choices of $\Omega_0$.}
\label{tab:delta_ti}
\centering
\begin{tabular}{lcccc}
& $\Omega_0=0.20$ & $\Omega_0=0.27$ & $\Omega_0=0.30$ & $\Omega_0=0.4$ \\
\hline
$\nu$ & $5.49$ & $5.51$ & $5.52$ & $5.53$ \\
$\gamma$ & $0.49$ & $0.51$ & $0.51$ & $0.52$ \\
$\theta$ & $0.89$ & $0.86$ & $0.85$ & $0.82$ \\
\hline
\end{tabular}
\end{table}

To compare the extracted profiles with our model, we need an explicit expression of the function $g_\rmn{NE}$. A possible expression is obtained via the spherical infall model, considering the evolution of a spherical perturbation from a primordal time $t_i$ to the present time $t_0$. According to \citet{Lal} and \citet{LL}, the overdensity profile of a primordial perturbation can be written in the following way:
\begin{eqnarray}\label{eq:delta_ti}
\delta(r,t_i)&=&\f{3}{2\pi^2\sigma_0 r}\f{D(t_i)}{D(t_0)}\int_0^\infty k j_1(kr) P(k)e^{-(R_f k)^2/2}\times\nonumber\\
&&\times\left[\f{\nu-\gamma^2\nu-\gamma\theta}{1-\gamma^2}+\f{\theta R_*^2}{3\gamma(1-\gamma^2)}k^2\right]dk.
\end{eqnarray}
In this equation, $P(k)$ is the power spectrum of the perturbation \citep{BBKS}, $D(t)$ is the growing solution of density fluctuations \citep{CPT}, $j_1$ is the first-order spherical Bessel function, $R_f$ is a filtering scale, $\sigma_0$ is the rms fluctuation of the filtered density field, and the parameters $R_*$, $\nu$, $\gamma$, and $\theta$ are related to the number density of peaks in the filtered density field \citep{LL}. We refer to above quoted papers for further details. All the parameters were tuned to match the conditions of the simulation we used. We chose in particular $\sigma_8=0.8$ and $R_f=0.1 h^{-1}$ Mpc, to avoid smoothing of fluctuations at the Mpc scale. The expectation values of $\nu$, $\gamma$, and $\theta$ were calculated in order to reproduce the number density of clusters in the simulation (which adopts $\Omega_0=0.3$); moreover, the values reported in Table \ref{tab:delta_ti} evidence the pure dependence on $\Omega_0$ of $\nu$, $\gamma$, and $\theta$.

Once the primordial overdensity profile is known, the spherical infall model provides a way to compute the corresponding present-day profile. The details are discussed in App. \ref{app:spherical_flat}. We obtain, in a comoving framework:
\be\label{eq:delta_evo}
1+\delta(r,t_0)=\left(\f{a_0}{r(t_0)}\right)^{-3}(1+\delta(r,t_i)),
\ee
where the ratio $r/a$ at any time is obtained by numerical integration of the Friedmann equation for the perturbation. The function $g_\rmn{NE}$ is thus determined as the best fit function for the present-day mass profile, according to equation (\ref{eq:M_NE}). We performed the computation for different values of $\Omega_0$ in the range $0.2\le\Omega_0\le 0.4$. Thus, the best fit function we adopted is:
\be
g_\rmn{NE}(r)=\exp\left[\f{K}{\Omega_0^{1/4}}\left(\f{r}{r_t}-1\right)\right]
\ee
where $K=0.6\pm 0.1$ (the uncertainty is due to the fitting algorithm). Once again the dependence on cosmology is very small. Assuming the concordance value $\Omega_0$ we obtain $g_\rmn{NE}(r)\simeq\exp[0.8(r/r_t-1)]$. This relation is adopted hereinafter to compute the theoretical profiles $\delta_\rmn{NE}(r)$ and $M_\rmn{NE}(r)$ from equation (\ref{eq:delta_NE}) and equation (\ref{eq:M_NE}). We can also compute the \emph{normalized} theoretical profiles as follows:
\be\label{eq:norm_i_delta}
\tilde\delta_\rmn{NE}(\tilde r)=\tilde r^{-3}\exp[0.8(\tilde r-1)],
\ee
\be\label{eq:norm_i_M}
\tilde M_\rmn{NE}(\tilde r)=\exp[0.8(\tilde r-1)].
\ee

The two plots of Fig. \ref{fig:prof_M} show the comparison between the normalized profiles extracted from the simulated cluster sample and the corresponding normalized profiles predicted by our spherical model. As one can see, the distribution of both $\tilde\delta_{i,j}^{(1)}$ and $\tilde M_{i,j}^{(1)}$ is in good agreement with $\tilde\delta_{\rmn{NE}}(\tilde r)$ and $\tilde M_{\rmn{NE}}(\tilde r)$ (solid lines). The simulated profiles show an intrinsic variance, due to the possible presence of different cluster substructure (filaments, clumps, or even a bimodal core). We can roughly distinguish two basic patterns: 
\begin{enumerate}
\p Regular profiles, i.e. smooth monotonic profiles, 
\p Irregular profiles, i.e. profiles showing one or more important changes of slope.
\end{enumerate}
A typical regular profile and two irregular profiles are shown in Fig. \ref{fig:prof_M}, with narrow solid lines and dashed lines, respectively. We qualitatively recognized in our sample $72$ regular profiles and $42$ irregular profiles (about $63\%$ and $37\%$, respectively). Usually, regular profiles are expected to produce the best agreement with our integrated model, while profiles with changes of slope are expected to deviate from our prediction. 

We tested the behaviour of the profiles by analysing the profiles of all the clusters of our catalogue when normalized by the turnaround radius and the turnaround overdensity (see points (1) and (2)). We say that a cluster (either regular or irregular) fits our model when its mass profile satisfies everywhere, in the considered region, the following condition ($i$ is the cluster label, and $j$ is the label of the cluster shell considered):
\be\label{eq:M_agr}
\left|\log_{10}\left[\f{\tilde M_{i,j}}{\tilde M_\rmn{NE}(\tilde r_{i,j})}\right]\right|\le\epsilon,\qquad 0.5\le\tilde r_{i,j}\le 2.
\ee
where $\epsilon$ defines the amplitude of the agreement band, and the $0.5\le\tilde r_{i,j}\le 2$ interval is defined in order to reproduce the non-equilibrium region; its amplitude is necessarily related to the amplitude of the non-equilibrium regions of our clusters. We first adopted the normalization in equation (\ref{eq:norm_i}), and introduced $\tilde r_{i,j}^{(1)}$ and $\tilde M_{i,j}^{(1)}$ into equation (\ref{eq:M_agr}). We performed three test for each cluster, with $\epsilon=0.10$, $\epsilon=0.15$, and $\epsilon=0.20$; these values correspond to a maximum ratio between data and model of about $1.25$, $1.4$, and $1.6$, respectively. The results of this comparison are reported in Table \ref{tab:type_agr}. For the three choices of $\epsilon$ (listed in columns), the first two rows indicate the number of regular and irregular clusters which agree with our model, respectively, while the third row indicates the \emph{total} number of clusters which agree with the model. Most clusters seem to be in good agreement with our model and therefore evidence the existence of a common profile in the external region. As expected, equation (\ref{eq:delta_NE}) and equation (\ref{eq:M_NE}) are particularly suitable to describe regular profiles, while poorly fit parts of the irregular profiles. In fact, in most of the cases of poor fit the discordance is present only in the extreme outskirts of the clusters, and does not bias the estimation of the overdensity and the mass at the turnaround radius.

\begin{table}
\caption{Agreement with the model for different types of mass profiles. We used the condition in equation (\ref{eq:M_agr}) with the normalization in equation (\ref{eq:norm_i}).}
\label{tab:type_agr}
\centering
\begin{tabular}{lccc}
& $\epsilon=0.10$ & $\epsilon=0.15$ & $\epsilon=0.20$ \\
\hline
Regular & $48/72$ ($67\%$) & $70/72$ ($97\%$) & $72/72$ ($100\%$) \\
Irregular & $19/42$ ($45\%$) & $29/42$ ($69\%$) & $38/42$ ($91\%$) \\
All & $67/114$ ($59\%$) & $99/114$ ($87\%$) & $110/114$ ($97\%$)\\
\hline
\end{tabular}
\end{table}

The normalization procedure used so far implies the knowledge of the infall velocity profile of clusters in order to compute $r_{t;i}$, $\delta_{t;i}$, and $M_{t;i}$. But the infall velocity profile cannot be computed directly from observations, since we know only the line-of-sight velocity component. So, to better estimate the model reliability when applied to observed clusters, we should adopt the normalization in equation (\ref{eq:norm_ii}). We therefore introduced $\tilde r_{i,j}^{(2)}$ and $\tilde M_{i,j}^{(2)}$ into equation (\ref{eq:M_agr}). The results of the comparison between the cluster profiles and those of our model are reported in Table \ref{tab:type_agr_2}. The overall agreement between the profiles is worse than that obtained via the previous normalization procedure; however, despite the variance among the infall velocity profiles of the clusters, about $80\%$ of all our profiles succeed to be well described by our model in all the non-equilibrium region (with a maximum uncertainty corresponding to $\epsilon=0.20$).

The second panel of Table \ref{tab:type_agr_2} displays the agreement between data and model at the turnaround radius $r_t$. In this case, we restricted the agreement interval in equation (\ref{eq:M_agr}) to $r_{ij}^{(2)}=r_{t,\rmn{RM}}$. As one can see, the difference between regular and irregular profiles is smaller than it is in the previous case. Despite the uncertainty due to the variance among clusters, our model is able to correctly estimate the cluster turnaround masses in more than $80\%$ of cases, within an agreement amplitude $\epsilon=0.15$.

\begin{table}
\caption{Agreement with the model for different types of mass profiles. We used the condition in equation (\ref{eq:M_agr}) with the normalization in equation (\ref{eq:norm_ii}).}
\label{tab:type_agr_2}
\centering
\begin{tabular}{lccc}
\multicolumn{4}{c}{\emph{Agreement in the whole non-equilibrium region}}\\
& $\epsilon=0.10$ & $\epsilon=0.15$ & $\epsilon=0.20$ \\
\hline
Regular & $23/72$ ($32\%$) & $48/72$ ($67\%$) & $64/72$ ($89\%$) \\
Irregular & $12/42$ ($29\%$) & $21/42$ ($50\%$) & $28/42$ ($67\%$) \\
All & $35/114$ ($31\%$) & $69/114$ ($61\%$) & $92/114$ ($81\%$)\\
\hline
\multicolumn{4}{c}{\emph{Agreement at turnaround radius $r_t$}}\\
& $\epsilon=0.10$ & $\epsilon=0.15$ & $\epsilon=0.20$ \\
\hline
Regular & $43/72$ ($60\%$) & $67/72$ ($93\%$) & $71/72$ ($99\%$) \\
Irregular & $16/42$ ($38\%$) & $25/42$ ($60\%$) & $32/42$ ($76\%$) \\
All & $59/114$ ($52\%$) & $92/114$ ($81\%$) & $103/114$ ($90\%$)\\
\hline
\end{tabular}
\end{table}
  
\section{Conclusions}\label{sec:concl}

We analized a large sample of simulated galaxy clusters in order to reconstruct the mass profile in the non-equilibrium region, where the galaxy dynamics is dominated by an overall infall motion towards the cluster centre. Within the assumptions of the spherical infall model, the turnaround overdensity $\delta_t$ can be theoretically computed as a function of only the matter density parameter $\Omega_0$, assuming a spatially flat universe. We obtained the overdensity 
$\delta_t\simeq 6-15$, depending on the infall velocity profile we adopted. 

We interpolated the infall velocity profile of member galaxies extracted from the simulated clusters of our catalogue, and we showed that:
\begin{enumerate}
\p The turnaround radius $r_t$ can be quite well approximated by a multiple of the virialization radius $r_v$: $r_t\simeq 3.5r_v$;
\p The turnaround overdensity $\delta_t$ is consistent with the prediction of the spherical infall model, as long as the infall velocity profile is described by the Meiksin approximation \citep{VD}.
\end{enumerate}
Points (i) and (ii) are in agreement with \citet{VH} and \citet{RG} and imply a proportionality between the turnaround mass $M_t$ and the virialization mass $M_v$: $M_t\simeq 1.7 M_v$. Moreover, $M_t$ turns out to depend on the 3-d DM velocity dispersion within the virialization core $\sigma_{v,\rmn{DM}}$ approximately in the form of a cubic relation. 

The turnaround values can be assumed as a suitable normalization scale for the mass profiles in the non-equilibrium region of clusters. We showed that the normalized mass profiles are generally consistent with a cosmic profile, which can be described (for $0.5\la r/r_t\la 2$) by: 
\be\label{eq:M_concl}
M(r)\simeq M_t\exp\left[\f{0.6}{\Omega_0^{1/4}}\left(\f{r}{r_t}-1\right)\right].
\ee
While in the inner, relaxed or almost-relaxed regions the mass can be considered independent on cosmological parameters, in the outer regions a dependence on $\Omega_0$ (even if small) has to be taken, at least in principle, into account.

We used a synthetic cluster, obtained by summing all catalogue clusters, to determine a robust estimate of $r_t$ and $\delta_t$. If we assume this values, our model is able to predict the mass profile in the non-equilibrium region for about $80\%$ of clusters. So, it is possible to speak about a mass profile even in the region where mass accretion takes places along isolated radial filaments rather than in a spherically symmetric way.

Our model may be useful in observational analysis in order to estimate the total mass of clusters using the redshift-space distribution of galaxies. The method is the following: 
\begin{enumerate}
\item one estimates the virialization radius and the virialization mass from the galaxy velocity dispersion in the cluster core;
\item using equation (\ref{eq:st}) and equation (\ref{eq:Mt}) one computes the turnaround radius $r_t$ and the turnaround mass of the cluster $M_t$;
\item once the turnaround radius and the turnaround mass are known, one can estimate the mass profile in the non-equilibrium region using the exponential law in equation (\ref{eq:M_concl}). 
\end{enumerate}
The advantage of this approach lies in the possibility to estimate the mass profiles up to the far outskirts of clusters, where the caustic pattern is not generally recognizable \citep{DG,D}; up to now, these cluster outer volumes have been usually neglected in the evaluation of cluster total masses.

Actually, the turnaround mass is a more exhaustive evaluation of the total mass of the cluster. The steps leading to it consist in the abovementioned points (i), (ii), and (iii), and are expected to be applied to observed clusters.

\section*{Acknowledgments}

We wish to thank Stefano Borgani for making available to us the numerical simulation, and Andrea Biviano and Marisa Girardi for providing the simulated galaxy catalogue, and all of them for the useful discussions. We are indebted to the anonymous Referee for the comments and the useful suggestions.

\appendix

\section{The spherical infall model in a flat universe}\label{app:spherical_flat}

In this appendix we briefly recall some results of the spherical infall model which are useful for our discussion. We consider a spherical density perturbation in a flat universe with cosmological constant ($\Omega_0+\Omega_\Lambda=1$) and describe it as a Friedmann universe on its own. Both the radius $R$ of the perturbation and the universal scale factor $a$ are normalized with $a_0$, the present-day scale factor, and treated as adimensional quantities. The redshift is therefore defined as $z=a^{-1}-1$.

Our first aim is to compute the radius and the overdensity of the perturbation at the present day as a function of the corresponding primordial values. We use for this purpose the Friedmann equation for the perturbation: 
\be\label{eq:Friedmann}
\f{\rmn{d}^2R}{\rmn{d}t^2}=-\f{4\pi G}{3}\rho R+\f{\Lambda c^2}{3}R.
\ee
Here $G$ is the gravitational constant, $\Lambda$ is the cosmological constant, and $\rho=\rho_\itl{bg}(1+\delta)$ is the density of the perturbation. Equation (\ref{eq:Friedmann}) has no analytical solution, but can be solved numerically between an initial time $t_i$ and the present time $t_0$, assuming the well-known Friedmann solution for $a$:
\be
a(t)=\left\{\f{\Omega_0}{1-\Omega_0}\sinh^2\left[\f{3}{2}(1-\Omega_0)^{1/2}H_0 t\right]\right\}^{1/3}.
\ee 
We will hereinafter adopt the subscript $\itl{in}$ and $0$ to denote the initial and the present value of quantities, respectively. We choose the initial time so as to obtain $R_\itl{in}\simeq a_\itl{in}$. In the matter-dominated era, we have $\rho_0 R_0^3\simeq\rho_\itl{in} a_\itl{in}^3$ and $a_\itl{in}^3=\rho_{\itl{bg},0}/\rho_{\itl{bg},\itl{in}}$. Equation (\ref{eq:Friedmann}) can therefore be rewritten as follows:
\be\label{eq:FDE}
\ddot F+2\f{\dot a}{a}\dot F+\left(\f{\ddot a}{a}+\Omega_0-1\right)F+\f{\Omega_0(1+\delta_\itl{in})}{2a^3}F^{-2}=0,
\ee
where $F\equiv R/a$, and the dots denote first- and second-order derivatives with respect to $\tau\equiv H_0 t$. We favour equation (\ref{eq:FDE}) because it gives stabler results when integrated by computational means. We have $F_\itl{in}\simeq 1$ and $F_0=R_0$, which yields, in a comoving framework: 
\be
r_0=F_0 r_\itl{in},
\ee
\be
1+\delta_0=F_0^{-3}(1+\delta_\itl{in}),
\ee
where $r$ is the comoving radius of the perturbation.

Our second aim is to determine an analytical expression for the turnaround radius and the turnaround overdensity of the perturbation at the present time. In this case, we study the variation of $R$ with respect to $a$, which can be expressed as follows \citep{P3}:
\be\label{eq:spherical_flat}
\left(\f{\rmn{d}R}{\rmn{d}a}\right)^2=\f{R^{-1}+\omega_0 R^2-\kappa}{a^{-1}+\omega_0 a^2}.
\ee
Here $\omega_0=\Omega_0^{-1}-1$, and $\kappa>0$ parametrizes the overdensity of the perturbation. When $\kappa$ is large enough, the perturbation expands until it reaches a maximum radius $R_\itl{max}$ and then recollapses. The value of $R_\itl{max}$ comes as a solution of the following third-degree algebraic equation \citep{ECF}:
\be
\omega_0 R_\itl{max}^3-\kappa R_\itl{max}+1=0,
\ee
where the condition $\kappa\le\left(9\omega_0/4\right)^{1/3}$ is required for positive real solutions to exist. If so, we obtain:
\be\label{eq:r_max}
R_\itl{max}(\kappa)=\left(\f{4\kappa}{3\omega_0}\right)^{1/2}\cos[\theta(\kappa)],
\ee
where
\be
\theta(\kappa)\equiv\f{1}{3}\left\{2\pi-\arccos\left[\left(\f{9\omega_0}{4\kappa^3}\right)^{1/2}\right]\right\}
\ee
with $\pi/2\le\theta(\kappa)\le2\pi/3$. The redshift of maximum amplitude $z_\itl{max}$ and the redshift of virialization $z_v$ of the perturbation can both be computed by numerical integration of equation (\ref{eq:spherical_flat}):
\be\label{eq:z_max}
z_\itl{max}(\kappa)=\f{\omega_0^{1/2}}{\sinh[\phi_\itl{max}(\kappa)]}-1,
\ee
\be\label{eq:z_v}
z_v(\kappa)=\f{\omega_0^{1/2}}{\sinh[\phi_v(\kappa)]}-1,
\ee
where
\be
\phi_\itl{max}(\kappa)\equiv\f{3}{2}\omega_0^{1/2}\int_0^{R_\itl{max}(\kappa)}{\left[\f{R}{\omega_0 R^3-\kappa+1}\right]^{1/2}\rmn{d}R},
\ee
and $\phi_v(\kappa)=2\phi_\itl{max}(\kappa)$, since the integral term must be taken twice to consider both the expansion phase and the collapse phase.

The present-day turnaround radius $r_t$ is defined as the radius of a perturbation which is now reaching its maximum amplitude. Conversely, the present-day virialization radius $r_v$ is defined as the radius of a perturbation which reached its maximum amplitude in the past and is now setting to equilibrium after collapse. Let $\kappa_t$ and $\kappa_v$ be the overdensity parameters of this two pertubations, respectively. We obtain:
\be
r_t=R_\itl{max}(\kappa_t),
\ee
\be
r_v\simeq\f{1-\eta_v/2}{2-\eta_v/2}R_\itl{max}(\kappa_v) 
\ee
where $\eta_v\equiv2\omega_0 R_\itl{max}(\kappa_v)^3$ \citep{Lal}.
Substituting $z_\itl{max}(\kappa_t)=0$ into equation (\ref{eq:z_max}) and $z_v(\kappa_v)=0$ into equation (\ref{eq:z_v}), we obtain $\phi_\itl{max}(\kappa_t)=\rmn{arcsinh}(\omega_0^{1/2})$ and $\phi_\itl{max}(\kappa_v)=\phi_v(\kappa_v)/2=\rmn{arcsinh}(\omega_0^{1/2})/2$. Since $R_\itl{max}(r_t)$ and $R_\itl{max}(r_v)$ are known from equation (\ref{eq:r_max}), we can use these relations to evaluate $\kappa_t$ and $\kappa_v$, and consequently $r_t$ and $r_v$. We obtain in particular, as an original result: 
\be
\f{r_t}{r_v}=\f{2-\omega_0 R_\itl{max}(\kappa_v)^3}{1-\omega_0 R_\itl{max}(\kappa_v)^3}\f{\kappa_t^{1/2}\cos\theta_t}{\kappa_v^{1/2}\cos\theta_v}
\ee
where $\theta_t\equiv\theta(\kappa_t)$ and $\theta_v\equiv\theta(\kappa_v)$. Since $1+\delta_t=r_t^{-3}$, we also obtain
\be
1+\delta_t=\left[\f{3\omega_0}{4\kappa_t\left(\cos\theta_t\right)^2}\right]^{3/2}.
\ee

\section{Inversion of the Yahil's formula}\label{app:Y_inv}

Let $f_\rmn{Y}$ be the Yahil approximation function, defined in equation (\ref{eq:Y}). We consider its first-degree Taylor series expansion around the point $\delta_0$:
\be\label{eq:Y_taylor}
f_\rmn{Y}(\delta)=f_1(\delta)+\mathcal{O}(\delta-\delta_0)^2,
\ee
\be
f_1(\delta)\equiv f_\rmn{Y}(\delta_0)+\left.\f{\rmn{d}f}{\rmn{d}\delta}\right|_{\delta=\delta_0}(\delta-\delta_0).
\ee
If the second- and higher-order terms are negligible, we can substitute equation (\ref{eq:Y_taylor}) into equation (\ref{eq:f}), thus obtaining a linear equation which provides $\delta$ as a function of $\delta_0$ and $\Omega_0^{-0.6}v_r/H_0 r$. Choosing $\delta_0=10$, the non-linear terms turn out to be negligible in the range where turnaround occurs:
\be
|\mathcal{O}(\delta-10)^2|\le 0.05f_\rmn{Y}(\delta),\qquad 7\le\delta\le 20.
\ee
In this case we can write:
\be
\delta_\rmn{Y}\equiv\delta(f_\rmn{Y})\simeq\f{66}{17}11^{1/4}\Omega_0^{-0.6}\f{v_r}{H_0 r}-\f{50}{17},
\ee
giving $\delta_{\rmn{Y},t}\simeq 11$ (when $v_r=H_0 r$) and $\Omega_0^{0.6}f_\rmn{Y}(\delta_{\rmn{Y},t})\simeq 0.998$, very close to unity.

\end{document}